# Improving Data Security in Infrastructure Networks Based on Unipath Routing

**Karnavel K[1], Sakthivel M[2], Karuppasamy L[3]**

[1,2,3]**Department of Computer Science and Engineering, Anand Institute of Higher Technology, Chennai, Tamil Nadu - 603103, India.**

## Abstract

An infrastructure network is a self-organizing network which uses Access Point (AP) of wireless links that connecting one node with another. These nodes can communicate without using ad hoc, instead these nodes form an arbitrary topology (BSS/ESS) in which these nodes play the role of routers. Though the efficiency of Infrastructure networks is high, they are highly vulnerable to security attacks. Detecting/Preventing these attacks over the network is highly challenging task. Many solutions are proposed to provide authentication, confidentiality, availability, secure routing and intrusion avoidance in infrastructure networks. Providing security in such dynamically changing networks is a hard task. Characteristic of infrastructure network should also be taken into consideration in order to design efficient solutions. In this study, we focus on efficiently increasing the flow transmission confidentiality in infrastructure networks based on multi-path routing. In order to increase confidentiality of transmitted data, we take advantage of the existence of multiple paths between nodes in an infrastructure network with the help of Access Point. In this approach the original data is split into package and are forwarded through access point. The encrypted packets are then forwarded in different disjoint paths that exist between sender and receiver. Even if an attacker succeeds to obtain one or more transmitted packets, the probability of reconstructing the original message is very low.

**Keywords:** *Infrastructure networks Security Confidentiality Unipath routing Basic Service Set (BSS) and Extended Service Set (ESS)*

## 1. Introduction

A wireless is a decentralized type of wireless network[1]. The network is ad hoc because it does not rely on a preexisting infrastructure, such as routers in wired networks or access points in managed (infrastructure) wireless networks. Instead, each node participates in routing by forwarding data for other nodes, so the determination of which nodes forward data is made dynamically on the basis of network connectivity. In addition to the classic routing, s can use flooding for forwarding the data. An typically refers to any set of networks where all devices have equal status on a network and are free to associate with any other device in link range. Often refers to a mode of operation of IEEE 802.11 wireless networks. Using cooperative wireless communications improves immunity to interference by having the destination node combine self-interference and other-node interference to improve decoding of the desired signal.

## 2. Literature Review

- Jalel Ben Othman (2009)-implemented enhancing data security in ad-hoc networks based on multipath routing. In this, the original message to secure is split into shares that are encrypted and combined then transmitted along different disjoined existing paths between sender and receiver. Even if an attacker succeeds to obtain one or more transmitted shares, the probability that the original message will be reconstituted is very low [1].

- Liang Zhou (January-2010)-scheduled security critical multimedia applications in heterogeneous networks. it takes into account application's timing and security requirements in addition to precedence constraints. As it finds resource allocations heuristically it maximizes the quality of security and the probability of meeting deadlines for all the multimedia applications running on heterogeneous networks. If the distortion model constructed is not accurate scalable graph based application becomes failure thereby decreasing the flexibility and efficiency [2].

- William R Claycomb(March-2010)-implemented a node level security policy in wireless networks. It reduces significant attacks at the node levels through distribution model algorithm and identity based cryptography. As sensor networks continues to be part of everyday life security of the network is critical to maintain [3].

- Frank A. Zdarsky(April-2010)- improved security in wireless mesh backhaul(WMB) architecture. It resolves the underlying security issues for the cases that standard security solutions do not exist. As everything is based on basic assumptions WMB architecture is still susceptible to security vulnerabilities [4].

- PavloBykovyy(November-2011)-proposed a specialized interface for detectors network of alarm system. It

provides a high level of security inside the system and the cost is low as it reduces the cabling required. If the traditional detectors used in this system becomes failure then the entire system cannot be protected [5].

- Hannes Holm (December-2011)-improved the performance of automated network vulnerabilities using remediated security issues. It is independent of the system credentials used by the system. However manual effort is needed to provide complete accuracy and it is prone to false alarms in the networks [6].

- Aura Reggiani(2012)-improved network resilience for transport security by considering methodological reflections. Through this resilience the system can absorb shocks without catastrophic changes in the functional organization and an effective tool in understanding the evolutionary paths of complex spatial networks. Yet the measure of resilience is still a critical issue as they still remain at a formal theoretical level [7].

## 3. Related Work

Recently, several researches interesting in infrastructure networks security aspects (like authentication, availability, secure routing, intrusion detection, etc) do exist. We can classify existing approaches into five principal categories:

- Trust Models
- Key Management Models
- Routing Protocols Security
- Intrusion Detection Systems
- Multipath protocols

    We mention below, some important proposals concerning each category:

### 3.1. Distributed trust model

The idea in [1] is based on the concept of trust. It adopts a decentralized management approach, generalizes the notion of rust, reduces ambiguity by using explicit trust statement and makes easier the exchange of trust-related information via Recommendation Protocol. The Recommendation Protocol is used to exchange trust information. Entities that are able to execute the Recommendation Protocol are called agents. Agents use trust categories to express trust towards other agents and store reputation records in their private databases to use them to generate recommendations to other agents. In this solution, memory requirements for storing reputations and the behavior of the Recommendation Protocol are issues that were not treated.

### 3.2. Key management Using Piconet

A Bluetooth network is called a small network or piconet. It has eight stations. One of which is called primary station and the rest of the stations are called secondary's. All secondary stations synchronize their clocks and hoping sequence with the primary. A piconet can have only one primary station and the communication between them is one-one or one-many. Although a piconet can have a maximum of seven secondaries an additional eight secondaries can be in parked state. A secondary is synchronized with the primary but they cannot take part in communication until it is moved from the parked state. [3]

### 3.3. Key agreement based password

The work developed in [8] draws up the scenario of a group wishing to provide a secured session in a conference room without supporting any infrastructure. The approach describes that these is a Weak Password that the entire group will share (for example by writing it on a board). Then, each member contributes to create a part of the session key using the weak password. This secured session key makes it possible to establish a secured channel without any centralized trust or infrastructure. This solution is adapted, to the case of conferences and meetings, where the number of nodes is small. It is rather strong solution since it does not have a strong shared key. But this model is not sufficient for more complicated environments. If we consider a group of people who do not know each other and want to communicate confidentially, this model becomes invalid. Another problem emerges if nodes are located in various places because the distribution of the Weak Password will not be possible any more.

### 3.4. Secure unicast communication in infrastructure network

Unicast is the most suitable model for reducing the incurring network load, when traffic needs to be securely delivered from a single authorized sender to a large group of valid receivers. Provision of security or unicast sessions is realized through enciphering the session traffic with cryptographic keys. All unicast members must hold valid keys in order to be able to decrypt the received information. The problem of distributing and updating the cryptographic keys to valid members (key management), adds storage, communication and computational overhead to the network management. Key updates are required either periodically or on-demand, to accommodate membership changes in multicast groups. Security should consume as minimal energy as possible in updating the keys. Authors in [7] present a cross-layer algorithm that considers the node transmission power (physical layer) and the unicast routing tree (network layer) in order to construct an energy efficient key distribution scheme (application layer). That means they introduce a cross-layer design approach for key management in wireless multicast that distributes cryptographic keys in an energy-efficient way. This solution is targeted towards a very special case (key management for unicast in infrastructure networks ensuring energy efficiency) and is complicated and supposes that the power consumption of computations is significantly reduced due to advances in silicon technologies, which is not true for all infrastructure network devices.

### 3.5. Secure routing protocol for mobile infrastructure networks

An important aspect of infrastructure network security is routing security. The Secure Routing Protocol (SRP) presented in [9,6,12,10] counters malicious behavior that targets the discovery of topological information. SRP provides correct routing information (factual, up-to-date, and authentic connectivity information regarding a pair of nodes that wish to communicate in a secure manner). SRP discovers one or more routes whose correctness can be verified. After verification, route request are propagated to the desired trusted destination. Route replies are returned strictly over the reversed route, as accumulated in the route request packet. There is an interaction of the protocol with the IP layer functionality. The reported path is the one placed in the reply packet by the destination, and the corresponding connectivity information is correct, since the reply was relayed along the reverse of the discovered route. They ensure that attackers cannot impersonate the destination and redirect data traffic, cannot respond with stale or corrupted routing information, are prevented from broadcasting forged control packets to prevent the later propagation of legitimate queries, and are unable to influence the topological knowledge of be being nodes.

### 3.6. Intrusion detection

One of recent interesting aspects of security in wireless networks, especially infrastructure networks is intrusion detection. It concerns detecting inappropriate, incorrect, or anomalous activity in the network. In [11] authors examine the vulnerabilities of wireless networks and argue that intrusion detection is a very important element in the security architecture for mobile computing environment. They developed such architecture and evaluated a key mechanism in this architecture, anomaly detection for mobile infrastructure network through simulation experiments. Intrusion prevention measures, such as encryption and authentication, can be used in infrastructure networks to reduce intrusion, but cannot eliminate them.

### 3.7. Shuffling

The security for reliable data delivery scheme addresses data confidentiality and data availability in a hostile ad hoc environment. The confidentiality and availability of messages exchanged between the source and destination nodes are statistically enhanced by the use of unipath routing. At the source, messages are split into multiple pieces that are sent out via single independent path. The destination node then combines the received pieces to reconstruct the original message. The Shuffle scheme used shuffling the frame in order of sequence before transmitting the data. In the continuation to avoid intruder problem in ad hoc network Fig.1.

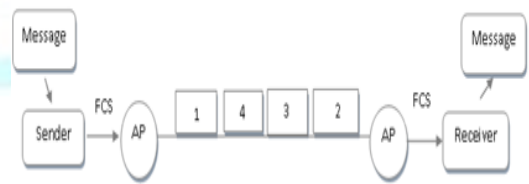

Fig: 1 Infrastructure unipath routing process

## 4. Securing data based unipath routing in s (SDMP)

The idea behind our protocol is to divide the initial message into parts then to encrypt and combine these parts by pairs. Then we exploit the characteristic of existence of single paths between nodes in an to increase the robustness of confidentiality. This is achieved by sending encrypted combinations on the same path between the sender and the receiver. In our solution, even if an attacker succeeds to have one part or more of transmitted parts, the probability that the original message can be reconstructed is low. We start by presenting our method principle.

## 4.1. Paths selection in SDMP

SDMP is based on unipath routing in infrastructure networks. The question is how to find the desired multiple paths in infrastructure network and how to deliver the different message parts to the destination using these paths? Routing in a infrastructure network presents great challenge because the nodes are capable of moving and the network topology (Fig.2 and Fig.3) can change continuously and unpredictably.

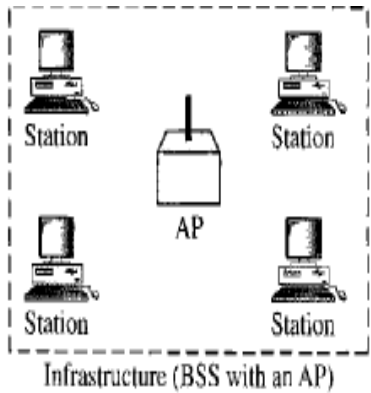

Fig: 2 Infrastructure network with BSS

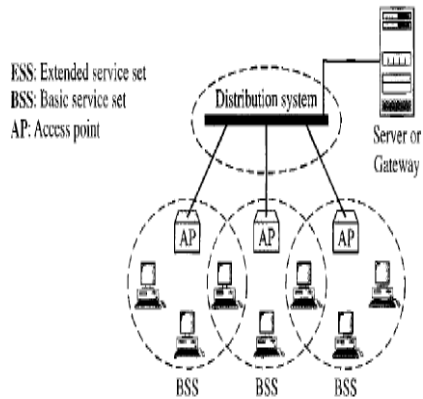

Fig: 3 Infrastructure network with ESS

## 5. Process Flowchart for collision Avoidance

The flowchart exploring the flow of process in wireless link, before exchange the frams and the distnace between stations can be great. Signal fading prevent a station at one end from hearing a collision at the other end Fig.3.

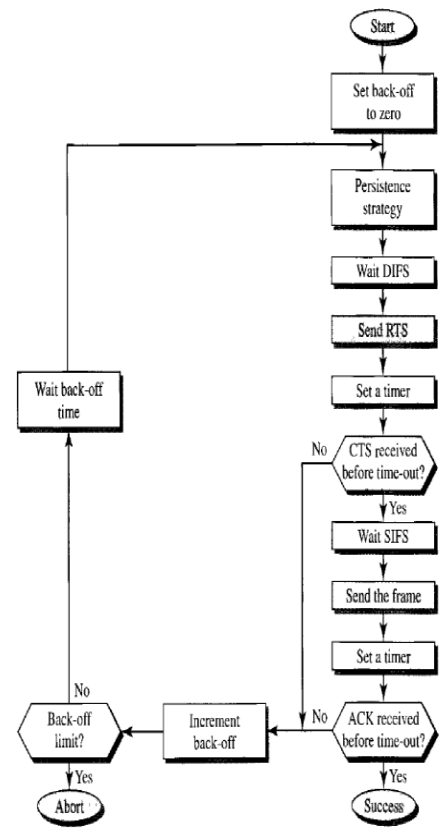

Fig: 4 Process Flowchart

## 6. Conclusion

In this paper, we proposed a new solution that treats data confidentiality problem by exploiting a very important infrastructure network feature, which is used the existence of unipaths between nodes. Our proposal improves data security efficiently without being time-consuming. It is not complicated and can be implemented in different infrastructure network. This protocol is strongly based on unicast routing characteristics of infrastructure networks and uses a route selection based on security costs and BSS, ESS provide the high data security for secure data transmission without using access point (AP). In this continuation our process before sending the data to check the path is idle or busy, afterword's, to send the data. SDMP protocol can be combined with other solutions which consider other security aspects than confidentiality to improve significantly the efficiency of security systems in infrastructure network.

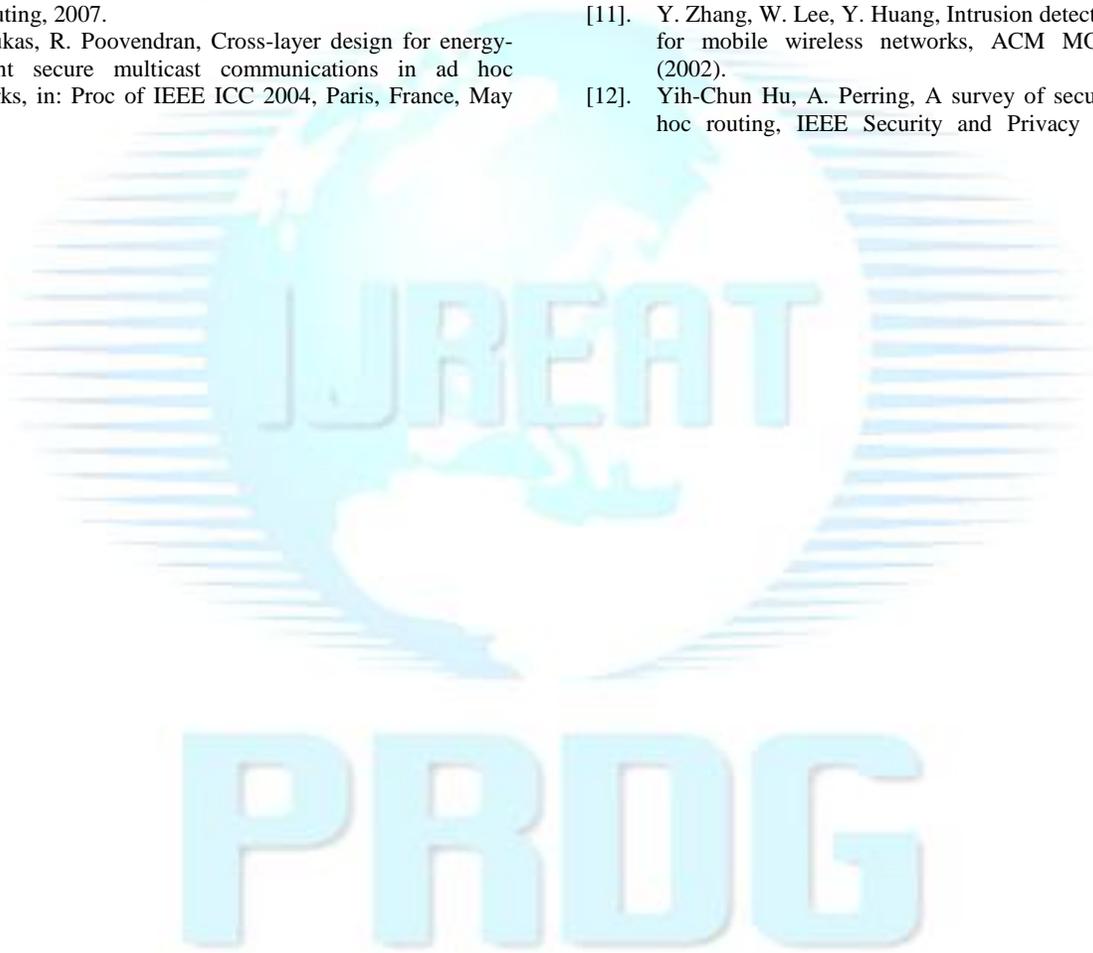